\documentclass[superscriptaddress,aps,twocolumn,prl,amsmath,amssymb]{revtex4-1}

\usepackage{amsmath}
\usepackage{amssymb}
\usepackage[varg]{txfonts}
\usepackage{color}
\usepackage[colorlinks]{hyperref}
\usepackage{tikz}
\usetikzlibrary{shapes}
\usepackage{xcolor}

\usepackage{subcaption}
\usepackage{graphicx}
\usepackage{bm}

\newcommand{\hexagon}{\mathord{\raisebox{0.6pt}{\tikz{\node[draw,scale=.65,regular polygon, regular polygon sides=6,fill=white](){};}}}}

\newcommand{\RNum}[1]{\uppercase\expandafter{\romannumeral #1\relax}}

\def\XXint#1#2#3{{\setbox0=\hbox{$#1{#2#3}{\int}$}
    \vcenter{\hbox{$#2#3$}}\kern-.5\wd0}}

\def\be{\begin{equation}}
\def\ee{\end{equation}}

\def\bi{\begin{itemize}}
    \def\ei{\end{itemize}}
\def\bn{\begin{enumerate}}
    \def\en{\end{enumerate}}
\def\bea{\begin{eqnarray}}
\def\eea{\end{eqnarray}}
\newcommand{\bpm}{\begin{pmatrix}}
    \newcommand{\epm}{\end{pmatrix}}

\def\ba{\begin{array}}
    \def\ea{\end{array}}
\def\bd{\begin{displaymath}}
\def\ed{\end{displaymath}}

\renewcommand{\imath}{\hspace{1pt}\mathrm{i}\hspace{1pt}}

\begin{document}

\title{Phase diagram of the Kitaev-Hubbard model: $\mathbb{Z}_2$ slave-spin and QMC approaches}

\author{Fatemeh Mohammadi}
\thanks{These two authors contributed equally to this work.}
\affiliation{Department of Physics, Sharif University of Technology, Tehran 14588-89694, Iran}

\author{S. Mojtaba Tabatabaei}
\thanks{These two authors contributed equally to this work.}
\affiliation{Department of Physics, Kharazmi University, Tehran 1571914911, Iran}

\author{Mehdi Kargarian}
\email{kargarian@sharif.edu}
\affiliation{Department of Physics, Sharif University of Technology, Tehran 14588-89694, Iran}

\author{Abolhassan Vaezi}
\email{vaezi@sharif.edu}
\affiliation{Department of Physics, Sharif University of Technology, Tehran 14588-89694, Iran}

\begin{abstract}
Recent experiments show that the ground state of some layered materials with localized moments is in close proximity to the Kitaev spin liquid, calling for a proper model to describe the measurements. The Kitaev-Hubbard (KHu) model is the minimal model that captures the essential ingredients of these systems; it yields the Kitaev-Heisenberg spin model at the strong coupling limit and contains the charge fluctuations present in these materials as well. Despite its relevance,  the phase diagram of the KHu model has not been rigorously revealed yet. In this work, we study the full phase diagram of the KHu model using the $\mathbb{Z}_2$ slave-spin mean-field theory as well as the auxiliary field quantum Monte Carlo on rather large systems and at low temperatures. The Mott transition is signaled by a vanishing quasiparticle weight evaluated using the slave-spin construction. Moreover, we demonstrate that there are multiple magnetic phase transitions within the Mott phase including magnetically ordered phases and most notably a quantum spin liquid phase for $1.0 \lesssim  t^{\prime}/t \lesssim 1.11 $ at $U/t=5$. 
\end{abstract}

\maketitle

\section {I. Introduction} 
Spin liquids form a class of topological phases in condensed matter physics that evade long-range magnetic orders even at absolute zero temperature \cite{Anderson1973, Anderson1987}. The protected ground state and the nontrivial statistics of excitations provide a unique platform for fault-tolerant quantum computations \cite{Nayak2008}. Magnetic insulators such as $\alpha$-RuCl$_3$ \cite{Banerjee2016, Kasahara:Nature2018, Banerjee2018} and iridium oxides \cite{Kim2009, Jackeli:PRL2009, Singh:PRL2012, Kitagawa:Nature2018}, with an underlying honeycomb lattice, are among the primary candidates for hosting spin liquid phases due to the interplay between electronic correlations and strong spin-orbit interactions. Despite the intense researches on these materials, all providing signatures of proximity to the Kitaev spin liquid, the unequivocal verification of a spin liquid state remains elusive. From the theoretical side, an effective spin model known as the Kitaev-Heisenberg model \cite{Jackeli:PRL2009, Chaloupka:PRL2010}, reminiscent of a strong interaction limit of the Hubbard model, has widely been studied. However, it turns out that adopting a correlated fermionic model may provide a better ground to explore the spin liquid phase in these materials; this is the main motivation of this work.

The fermionic Kitaev-Hubbard (KHu) model (see Eq.~\ref{H_KH}) was first introduced in Ref.\cite{Duan2003}, where the authors proposed it to simulate the Kitaev model in optical lattices. A three-orbital fermionic model, which preserves time-reversal symmetry, can also effectively represent KH model in strong coupling limit \cite{Rachel:PRB2017, Rachel2018}. Although the existence of a spin liquid phase in the KHu model has been investigated numerically \cite{Hassan:PRL2013, Hassan:PRB2014, Liang:PRB2014,Dong2023}, the presence of topological order, the Mott phase transition, and gapless fractionalized excitations in this fermionic model still need to be firmly established. For instance, the exact boundaries of the quantum spin liquid (QSL) phase remains unsettled, direct measurement of the spin excitation gap within the QSL is lacking, and the order of the phase transitions into the QSL is still debated. Furthermore, a systematic study of the quantum phase transitions upon tuning the spin-orbit coupling strength,  $t^{\prime}$, has yet to be thoroughly investigated.

In this letter, we examine the KHu model using three complementary methods: (i) the $Z_2$ slave-spin approach, (ii) the variational Monte Carlo (VMC) algorithm, and (iii) the determinant quantum Monte Carlo (QMC) algorithm. The $Z_2$ slave-spin method \cite{Huber:PRL2009, Ruegg:PRB2010, Fiete:PRL2012} provides a natural framework for studying the Mott transition. The VMC method \cite{Ceperley_prb_1977,Umrigar_prl_1988,Drummond_prb_2005,vaezi2018unified}, on the other hand, not only detects the Mott transition but also identifies various magnetic orders within the Mott insulating phase. Finally, the QMC approach \cite{white1989numerical,assaad2013pinning, vaezi2021amelioration,Tabatabaei_prl_2022} offers an unbiased, nonperturbative description of the system across a wide range of spin-orbit coupling strengths \cite{assaad2013pinning}. In particular, by employing the QMC algorithm, we are able to calculate both the static and dynamical susceptibilities in the spin and charge sectors, as well as reliably measure the spin and charge excitation gaps. Furthermore, this method allows us to evaluate the first and second derivatives of the ground-state energy with respect to the tuning parameters, and to compute the correlation ratio, which helps identify quantum critical points (QCPs). Using these tools, we confirm the emergence of the Kitaev quantum spin liquid for $1.0 \lesssim  t^{\prime}/t \lesssim 1.11 $ at $U/t=5$, nestled between the antiferromagnetic (AF) and zigzag (ZZ) orders--echoing the insights drawn from Dong et al.'s recent fPEPS study \cite{Dong2023}.

\begin{figure}
\begin{centering}
\includegraphics[width=8.5cm]{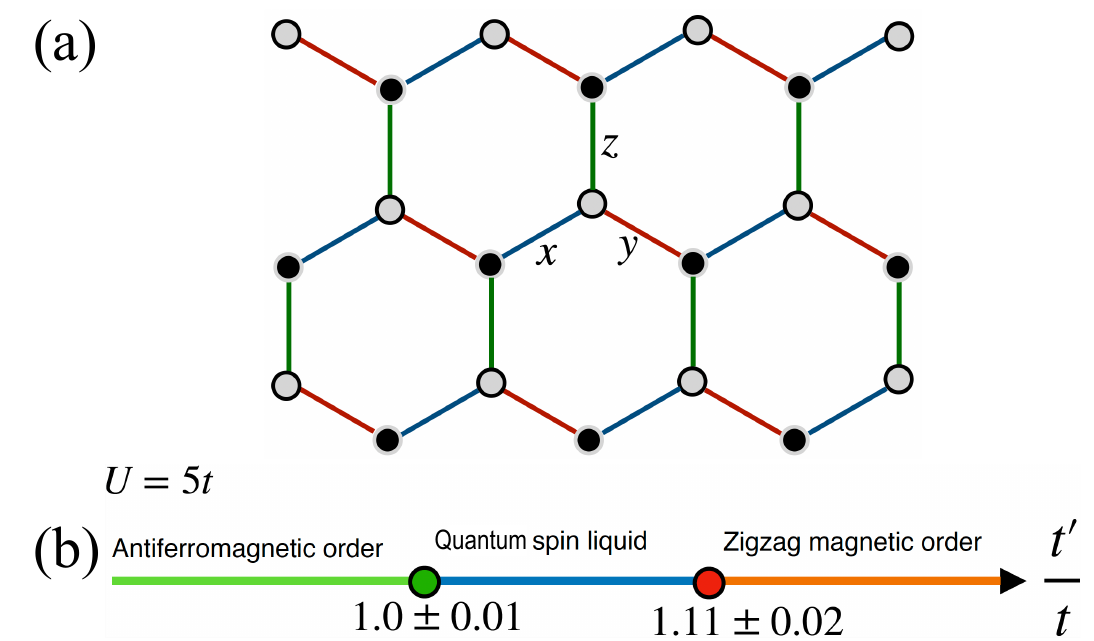}
\par\end{centering}
\caption{Lattice structure associated with the Kitaev-Hubbard model (a) and its phase diagram at $U/t=5$ as suggested by the QMC approach (b). \label{fig:QMC_phase_diagram}}
\end{figure}

\section{II. Model}
The KHu model \cite{Duan2003}, is described by the following spin-dependent Hamiltonian
\begin{equation}\label{H_KH}
H_{\mathrm{KHu}} =  - \sum_{\langle ij \rangle_\gamma} \left \lbrace c_i^{\dagger}\frac{(t+t^{\prime}\sigma_{\gamma})}{2}c_j + \mathrm{h.c.} \right \rbrace + U \sum_j \left(n_{j,\uparrow}-\frac{1}{2}\right)\left(n_{j,\downarrow}-\frac{1}{2}\right),
\end{equation}
where, $c_{i} = \left(c_{i,\uparrow},c_{i,\downarrow}\right)^{\rm T}$, and $c_{i,s}$($c_{i,s}^{\dagger}$) denotes the fermionic annihilation (creation) operator with spin $s$. 
The parameter $t$ represents a regular spin-independent hopping parameter, whereas $t^{\prime}\sigma_{\gamma}$ introduces anisotropy into the system through spin-dependent hopping along three distinct directions of nearest neighbors, where $\sigma_{\gamma}$ represents the Pauli matrix associated with the Cartesian coordinate $\gamma=(x,y,z)$. The system is subject to an onsite Coulomb repulsion $U$ and lives at half-filling. We set $t=1$ throughout this letter. The main interest in the KHu model is due to the fact that it yields the following Kitaev-Heisenberg effective spin model in $U\gg t,t^{\prime}$ limit~\cite{Duan2003,Hassan:PRB2014,sato2021quantum}
\begin{equation}\label{H_JK}
H_{\mathrm{KH}}=J\sum_{\langle ij \rangle}S_{i}.S_{j}+2K\sum_{\langle ij \rangle_\gamma}S_{i}^{\gamma}S_{j}^{\gamma},
\end{equation}
where, $K=A\sin\phi$ and $J=A\cos\phi$ with $A=\sqrt{(1-2t^{\prime2}+2t^{\prime4})}/U$
and $\tan\phi=t^{\prime2}/(1-t^{\prime2})$. Here, $t^{\prime}=1$ corresponds to the pure Kitaev model. 

\section {III. Slave-spin approach} 
Let us first develop an analytical understanding of the Mott transition in the KHu model. To this end, we employ the $\mathbb{Z}_2$ slave-spin approach where, unlike other slave-spin methods~\cite{Kotliar:PRL1986, Florens:PRB2002, Florens:PRB2004, deMedici:PRB2005,mardani2011slave,HassanPRB2010}, involves a discrete $\mathbb{Z}_2$ gauge fluctuations and extend the local Hilbert space with $\rm dim = 4$ into a space with $\rm dim = 8$. Within this method, each fermion is represented using an auxiliary spin as follows
\begin{equation}\label{slavespin}
c_{i,s}^{(\dagger)}=2I^x_i f_{i,s}^{(\dagger)},
\end{equation} 
where, $I_i^x$ is a pseudo-spin carrying the electron charge property, and $f_{i,s}$ is the annihilation operator for a pseudo-fermion (see the Appendix B).  In this representation the KHu model Eq.~\eqref{H_KH} is given by the following Hamiltonian
\begin{equation}\label{KHM_slave_spin}
H_{\rm SS} = 4 \sum_{\langle ij \rangle_\gamma} I_i^x I_j^x f_{i,\alpha}^{\dagger}t_{\alpha,\beta}^{\gamma}f_{j,\beta} + \frac{U}{2} \sum_j I_j^z,
\end{equation}
where, the Hubbard interaction term has been transformed into a simple form involving the field $I_z$. However, the hopping term becomes quartic, which can be further simplified using the mean-field approximation. This facilitates an exact examination of the Hubbard term and an approximate analysis of the hopping term using mean-field theory. 
At the mean-field level, the above Hamiltonian can be decoupled into a fermionic and a quantum transverse field Ising Hamiltonians upon substituting pseudo-fermion and Ising bilinear terms with their expectation values 
\begin{equation}\label{mean_fields}
\begin{split}
&g_{i,j}=4\langle I_i^x I_j^x\rangle, \quad J_{i,j}=-\frac{1}{2}\langle f_{i,s}^{\dagger}\left(t\delta_{s,s'} + t^{\prime}\sigma_{s,s'}^\gamma\right)  f_{j,s'}^{\dagger} \rangle + \text{c.c.}
\end{split}
\end{equation}

The resulting Ising model is subject to an in-plane magnetic field of strength $U/2$ and a nearest neighbor coupling of strength $4\left|J_{\gamma}\left(t^{\prime}\right)\right|$ and illustrates two distinct phases in $2+1$D. Upon increasing the transverse field strength ($U$) beyond a critical value $U_c\left(t^{\prime}\right)$, the system undergoes a quantum phase transition from a ferromagnetic to a paramagnetic state. The paramagnetic phase is characterized by $\langle I_i^x\rangle=0$. Since $I_x$ encapsulates the charge degrees of freedom by construction, in this phase the charge fluctuations are suppressed. Thus, the Mott phase is characterized by vanishing the quasi-particle weight
$Z = 4 \langle I_i^x \rangle^2$ \cite{Florens2004, HassanPRB2010}. In Appendix B, we present our cluster mean-field approximation to calculate $g_{\gamma}$ parameters. 
Our numerical analyses indicate the model is a semimetal (SM) in the limit of weak coupling and  a Mott phase arises for large values of $U$. Moreover, since $\left|J_{\gamma}\left(t^{\prime}\right)\right|$ increases with $t^{\prime}$ away from $t^{\prime} = 0$,  the Ising model's nearest neighbor coupling becomes stronger and thus a higher value of the in-plane magnetic field strength , i.e., $U$, is required for the Ising (Mott) transition to occur.

The $\mathbb{Z}_2$ slave-spin representation of the KHu model yields a topological $\mathbb{Z}_2$ gauge description \cite{Senthil:PRB2000} of low-energy excitations. The slave-spin construction in Eq.~\eqref{slavespin} has a local $\mathbb{Z}_2$ symmetry: $I^x_{i}\rightarrow \epsilon_i I^x_{i} $ and $f_{i,s}\rightarrow \epsilon_i f_{i,s}$ with $\epsilon_i=\pm$ which leaves $c_{i,s}$ invariant. The mean-field Hamiltonian, however, breaks the $\mathbb{Z}_2$ symmetry. To restore the symmetry and go beyond the mean-field limit, we introduce $\mathbb{Z}_2$ gauge fields $\sigma^z_{ij}$ on the spatial links. The low-energy gauge theory (see Appendix C) reads as \cite{Fiete:PRL2012,Maciejko2014} 
\begin{equation}
H_{\sigma} = -K  \sum_{\hexagon}\prod_{\langle ij\rangle \in \hexagon}\sigma_{ij}^z - G \sum_i \prod_{j(i)}\sigma_{ij}^x - I \sum_{\langle ij \rangle} \sigma_{ij}^x,
\end{equation}
where $K$, $G$ and I are some constants. This model possesses a confined-deconfined phase transition, where the deconfined phase corresponds to a topological spin-liquid phase in the original lattice Hamiltonian. 

\begin{figure}
\begin{centering}
\includegraphics[width=8.5cm]{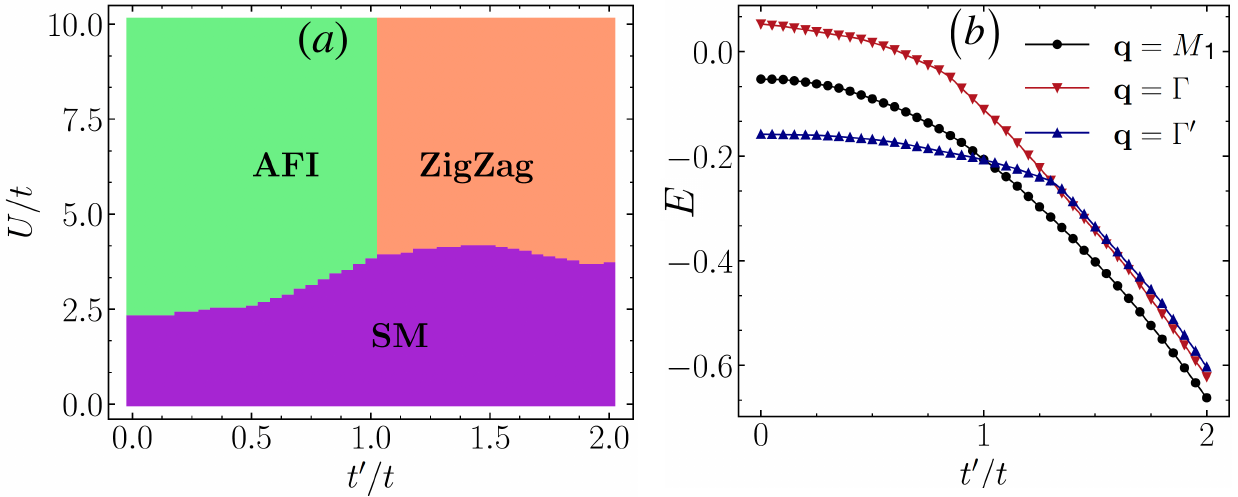}
\par\end{centering}
\caption{(a) VMC's prediction for the phase diagram of the Kitaev-Hubbard model. (b) The energy comparison of three putative projected spin wave configurations at $U/t=5$. 
\label{fig:VMC_phase_diagram}}
\end{figure}

\section{IV. Variational Monte Carlo results}
To gain further insights, let us now apply the VMC approach \cite{Ceperley_prb_1977,Umrigar_prl_1988,Drummond_prb_2005,vaezi2018unified} to the KHu model for a lattice with $L\times L$ unit cells ($2L^{2}$ lattice points) up to $L=18$. In particular, we are interested in identifying the phase boundary between the Mott insulator and the SM phases $U_c(t^{\prime})$, as well as various magnetic phases within the Mott insulator.  To this end, we consider a family of Gutzwiller projected states with $\left| \Psi_{\lambda_{\rm GW}, r, \bf m, q}\right \rangle_{\rm VMC} \propto \prod_{i} e^{-\lambda_{\rm GW} n_{i,\uparrow}n_{i,\downarrow}}\left| \Phi_{r, \bf m, q}\right \rangle_{\rm MF}$ structure, where $\lambda_{\rm GW}$ is the Gutzwiller variational parameter and $\left| \Phi_{r, \bf m, q}\right \rangle_{\rm MF}$ is the ground state of the following spin-density-wave mean-field Hamiltonian 
\begin{equation}\label{MFH1}
H_{\rm MF} =  - \sum_{\langle ij \rangle_\gamma} \left \lbrace c_i^{\dagger}\frac{(t+ r t^{\prime}\sigma_{\gamma})}{2}c_j + \mathrm{h.c.} \right \rbrace - \sum_{i} \cos\left({\bf q.r}_i\right) {\bf m. s}_i ,
\end{equation}
where $s_{i,a=x,y,z} = c_{i,\alpha}^\dag \sigma^{a}_{\alpha,\beta} c_{i,\beta}$ are spin operator components. Here, we treat $r$, $m_{x,y,z}$, and $q_{x,y}$ as six additional variational parameters describing the renormalized $t^{\prime}$, and the amplitude and wave-vector of the magnetic ordering, respectively. By minimizing the expectation value of our model Hamiltonian $H_{\mathrm{KHu}}$ w.r.t. $\left| \Psi_{\lambda_{\rm GW}, r, \bf m, q}\right \rangle_{\rm VMC}$, the optimum values of $\lambda_{\rm GW}$, $r$, $\bf m$ and $\bf q$ can be obtained (see Fig.~\ref{fig:VMC_phase_diagram}). To this end, we utilized the Nelder-Mead approach which is gradient free and can achieve the optimum values efficiently.

For weak onsite Hubbard interaction, the optimal energy occurs when $\bf{m} = 0$, indicating that no magnetic ordering is expected, corresponding to the semimetal (SM) phase. As $U$ increases, for $t^{\prime}/t < 1$, the energy is minimized with a non-vanishing $\bf m$ and ${\bf q} = \Gamma^{\prime}$ ($\in$ the $2$nd Brillouin zone), which corresponds to an AF insulator phase. Moving along the $t^{\prime}/t$ axis (above the Mott transition line), the system enters the ZZ phase with (${\bf q =} M_1$) for $t^{\prime}/t >1$. As illustrated in Fig.~\ref{fig:VMC_phase_diagram}(b), our VMC approximation indicates a first-order phase transition between the AF and ZZ phases, signaled by a discontinuous $\frac{dE}{dt^{\prime}}$ at $t^{\prime}/t=1$. In addition to the slave-spin and VMC approaches, we also applied the random phase approximation to study magnetic instabilities in the KHu model, obtaining similar results (see the Appendix B).

\section{V. Quantum Monte Carlo Calculations}
We observed that the aforementioned family of Gutzwiller wavefunctions is unable to capture the QSL phase, indicating the need for a more rigorous treatment of quantum fluctuations. To address this limitation and go beyond the VMC approximation, we now employ the QMC method to treat the model Hamiltonian in Eq.~\eqref{H_KH} exactly~\cite{white1989numerical,assaad2013pinning, vaezi2021amelioration}.

\begin{figure}
\begin{centering}
\includegraphics[width=5.0cm]{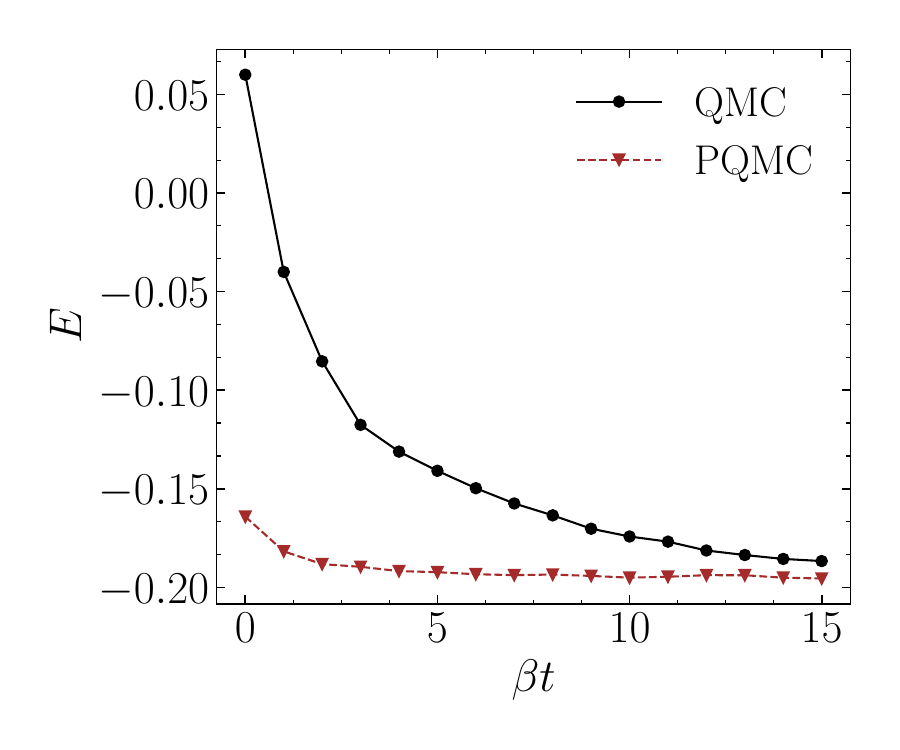}
\par\end{centering}
\caption{A comparison between the performance of QMC and PQMC approaches for $t^{\prime}=0.4t$ and $U/t=5$. 
\label{fig:E_beta}}
\end{figure}

\begin{figure}
\begin{centering}
\includegraphics[width=7.5cm]{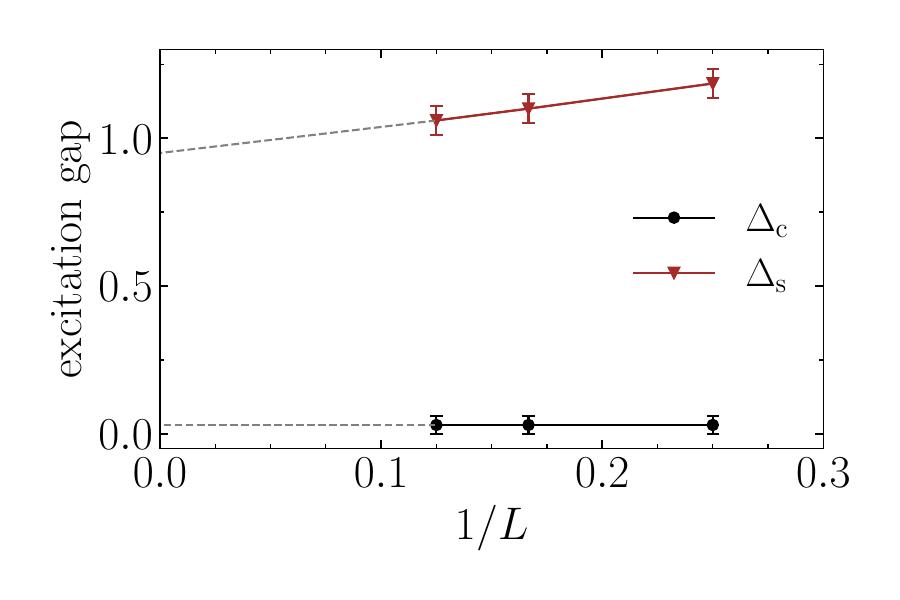}
\par\end{centering}
\caption{Finite size scaling of the spin ($\Delta_s$) and charge ($\Delta_c$) excitation gaps in the QSL phase for $t^{\prime}/t=1.05$ and $U/t=5$. Extrapolation to $L=\infty$ suggests $\Delta_s \to 0 $ and $\Delta_c \to 0.95$ using extrapolations of QMC to $\beta t \to \infty$  \label{fig:gaps}}
\end{figure}

To this end, we employ two families of the QMC algorithms: (i) The finite temperature auxiliary field QMC at $T=1/\beta$. (ii) The projector (ground-state) QMC (PQMC) with a trial wavefunction suggested by our VMC approximation above and then evolved in imaginary time using the full model Hamiltonian, i.e., $\left| \Psi_{\lambda_{\rm GW}, r, \bf m, q}\right \rangle_{\rm PQMC} \propto e^{-\tau H_{\mathrm{KH}}}\left| \Psi_{\lambda_{\rm GW}, r, \bf m, q}\right \rangle_{\rm VMC}$, where $\tau = \beta/2$ is the duration of the imaginary time evolution~\cite{Tabatabaei_prl_2022}. PQMC states describe pure (entropyless) states, where thermal fluctuations are absent, hence $\left| \Psi_{\lambda, r, \bf m, Q}\right \rangle_{\rm PQMC}$ better captures the zero temperature limit. As. Fig.~\ref{fig:E_beta} suggests, PQMC  converges to the true ground-state for relatively short evolution times thanks to its pure nature and optimized trial state. Our focus in this section will be on $U/t=5$ to investigate the claims that a quantum spin liquid is present for $t^{\prime}/t$ close to 1 around $U/t=5$~\cite{Hassan:PRL2013, Liang:PRB2014, Dong2023, sato2021quantum}. Our VMC results indicates that $U/t=5$ belongs to the Mott insulator phase for the entire parameter regime of interest $\left|t^{\prime}\right| < 1.35$. This was further confirmed by measuring the charge excitation gap in the QMC framework (see Fig.~\ref{fig:gaps} for one example).

Below, we consider $\beta t=2, 3,\cdots,12$ (and extrapolate to $\beta t=\infty$) and study $L=4,6,8$. Although, the KHu model suffers from the fermionic sign problem and its average sign diminishes quickly upon increasing $t^{\prime}$ beyond unity, we used increased samplings and thousands of independent Markov chains to keep the statistical error bars negligible and circumvent the fermionic sign problem to achieve virtually exact results. Furthermore, utilizing the adiabatic QMC algorithm~\cite{vaezi2021amelioration} can enhance the average sign and reduce the statistical noise significantly.

\begin{figure}
\begin{centering}
\includegraphics[width=8.9cm]{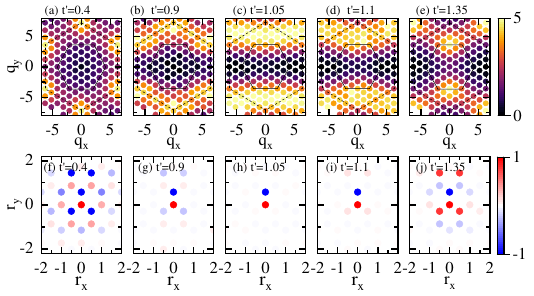}
\par\end{centering}
\caption{\label{fig:S6} Equal time spin-z correlation function shown in reciprocal (first row) and real (second row) space for a lattice with $L=6$ vs $t^\prime$ at $U/t=5$ using QMC simulations at $\beta=10$.}
\end{figure}

\begin{figure}
\begin{centering}
\includegraphics[width=8.9cm]{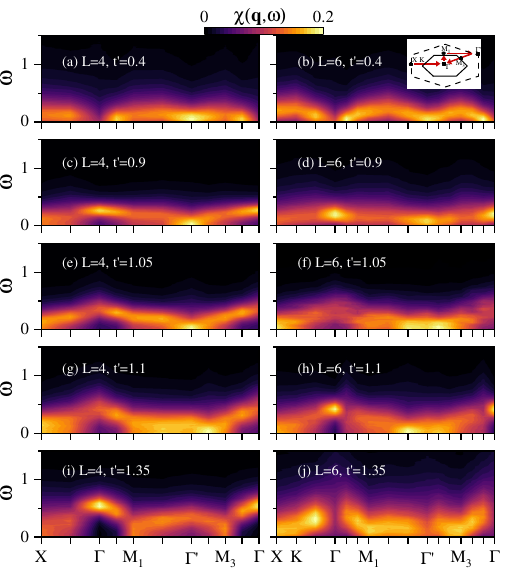}
\par\end{centering}
\caption{\label{fig:spec} Dynamical spin-spin susceptibilities for different values of $t^\prime$ for $L=4$ and $L=6$ ($U/t=5$) in the left and right panels, respectively, using QMC simulations at $\beta=10$.
}
\end{figure}

\begin{figure}
\begin{centering}
\includegraphics[width=8.9cm]{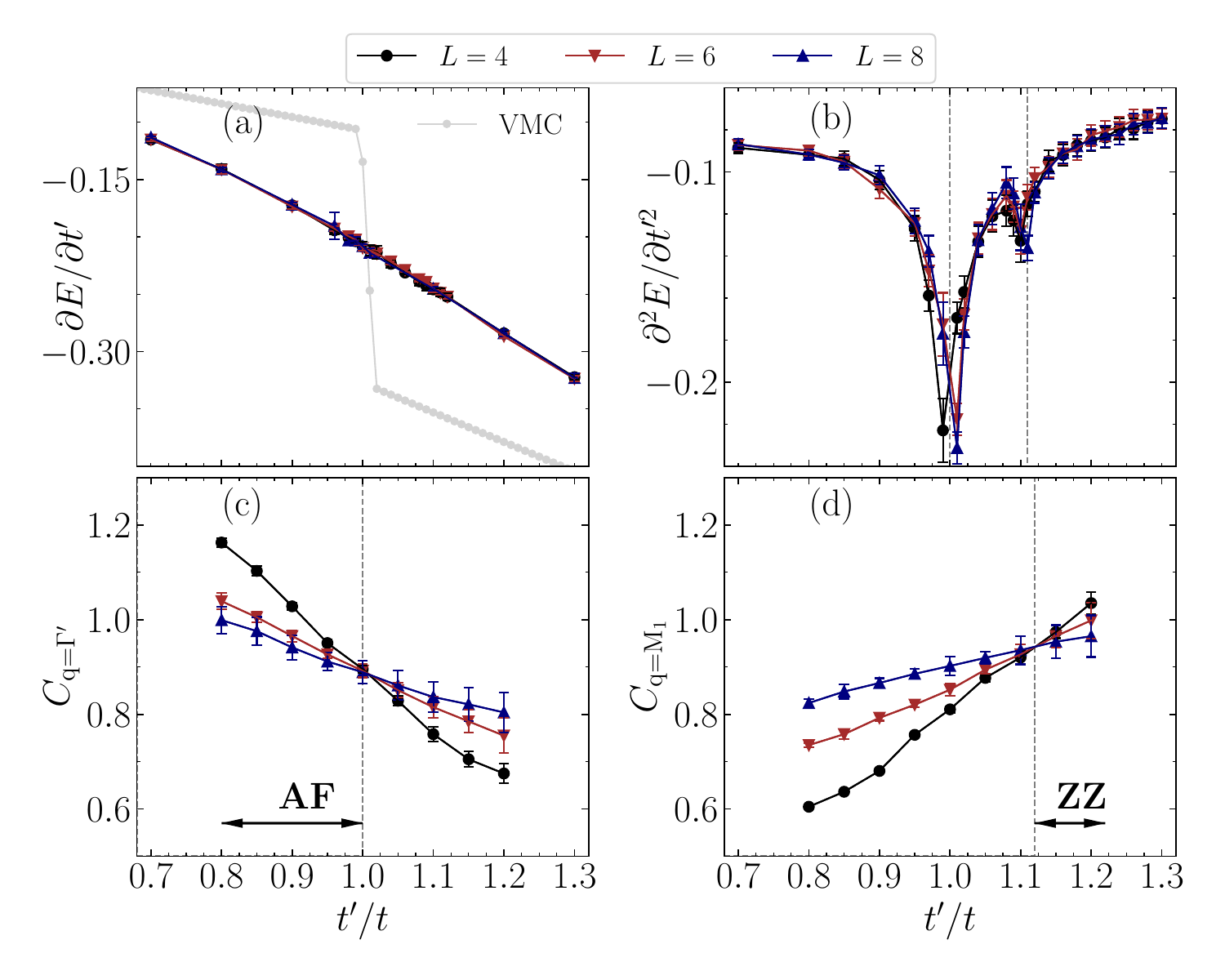}
\par\end{centering}
\caption{\label{fig:derivative_corr} (a)-(b) The first and second derivatives of the energy density w.r.t. $t^{\prime}$ at $U/t=5$ using the QMC. Unlike the VMC approach (the gray profile in panel (a)), which suggests a single first-order phase transition, our QMC results reveal two continuous transitions between the AF, QSL, and ZZ phases, as evidenced by the two peaks in the top right panel. (c)-(d) The crossing points of $C_{ \Gamma^{\prime}}(t^{\prime},L)$ and $C_{M_1}(t^{\prime},L)$ profiles locate the two quantum critical points. These results are obtained using the finite temperature QMC algorithm via extrapolation to $\beta t=\infty$ from $\beta t=4, \cdots, 12$ data points. We observed similar results using the PQMC via extrapolation to $\bf m \to 0$ and $\beta t \to \infty$.}
\end{figure}

Let us first study the dynamical spin-spin susceptibility defined as
\begin{equation}
\chi({\bf q},\omega)=\frac{i}{3}\sum_{\alpha=x,y,z}\frac{1}{L^2}\sum_{m,n}e^{i{\bf q}.({\bf r}_{m}-{\bf r}_{n})}\int_{0}^{\infty}dte^{i\omega t}\left\langle \left[S_{m}^{\alpha}(t),S_{{n}}^{\alpha}\right]\right\rangle ,
\end{equation}
where $m,n$ runs over all lattice points. The dynamical spin-spin susceptibility can teach us useful information about the spin excitation gap at different momenta and possible magnetic orderings. 
In Figs.~\ref{fig:spec} and \ref{fig:S6}, we present our QMC results
for $L=4$ ($32$ site) and $L=6$ ($72$ sites) in the right and left columns, respectively.
For small values of $t^{\prime},$ the system is in the AF Mott phase
where the spin susceptibility is gapless around $\Gamma^{\prime}$
point as it is seen in Figs.~\ref{fig:S6}(a,b).
This behavior is also obvious in the static spin susceptibilities
shown in Fig.~\ref{fig:S6}(a) where the maximum
value of the static susceptibility happens at $\Gamma^{\prime}$. Moreover, the real space correlation function in the $z$ direction
in Fig.~\ref{fig:S6}(f) also shows AF ordering.
On the other hand, the dynamical spin susceptibilities shown in Fig.~\ref{fig:spec} for
$t^{\prime}/t = 0.9, 1.05, 1.1$ exhibit a broad spectral gap
between $\Gamma$ and $M_{1}$  points. This is indicative of the system being in close proximity to a spin liquid phase. The corresponding static susceptibilities and real space configurations are shown in Fig.~\ref{fig:S6}(b-d) and (g-i), respectively, where it is clearly seen that the spin-spin correlations fall off rapidly with distance. For $t^{\prime}/t=1.35$, the correlations become long ranged as shown in Fig.~\ref{fig:S6}(e,j). Here, the system is
in the ZZ order where the spectra is mainly gapless around the $M_1$ and $M_3$ points (see Fig.~\ref{fig:spec}(i,j)).

Next, in order to identify the boundary between the AF, QSL and ZZ phases, we employ two further diagnostics: (i) We look into the first and the second order derivatives of the energy density w.r.t. the tuning parameter, $t^{\prime}$ which can be evaluated using the Hellmann-Feynman theorem. As Fig.~\ref{fig:derivative_corr}(a) shows, the first derivative remains continuous and the second order derivative in Fig.~\ref{fig:derivative_corr}(b) shows two peaks at $t^{\prime}/t=1\pm 0.01$ and $t^{\prime}/t=1.11 \pm 0.01$ respectively. While the two regions with $t^{\prime}/t \lesssim 1$ and $t^{\prime}/t \gtrsim 1.11$ are clearly associated with the AF and ZZ orders, the intervening region corresponds to the QSL phase as it does not show any magnetic order. To further corroborate this observation we study the correlation ratio defined as $C_{{\bf q}}={\chi_{z}({\bf q})}/{\chi_{z}({\bf q}^{*})},$
where ${\bf q}^{*}$ is the nearest point to ${\bf q}$ in the
reciprocal lattice.
The scaling analysis indicates that $C_{{\bf q=q_0}}$ is finite and independent of $L$ at the QCP ($t^{\prime}_c$), where the magnetic ordering associated with $\bf q=q_0$ starts to disappear. Thus, the crossing of $C_{{\bf q=q_0}}\left(t^{\prime},L\right)$ profiles identifies $t^{\prime}_c$. For our model, the two relevant magnetic orderings are the AF and ZZ orders (corresponding to ${\bf q} = \Gamma^{\prime}$ and $M_1$ wave-vectors, respectively). Figs.~\ref{fig:derivative_corr}(c,d) show that $C_{ \Gamma^{\prime}}\left(t^{\prime}, L\right)$ profiles cross at $t^{\prime}/t \approx 1$, hence for $t^{\prime}/t \gtrsim 1$, the AF order is killed. Moreover, $C_{ M_1}\left(t^{\prime}, L\right)$ profiles cross at $t^{\prime}/t\approx 1.12$. Thus, the ZZ order is absent for $t^{\prime}/t \lesssim 1.12$. The intervening region $1 \lesssim t^{\prime}/t \lesssim 1.12$ exhibits no magnetic ordering and therefore is a quantum spin liquid. To uncover the nature of this intermediate spin liquid region further, we measure the spin excitation gap vs $L$ for $t^{\prime}/t=1.05$ and extrapolate it to $L \to \infty$. These quantities can be extracted via an exponential fit to the decaying tail of the unequal time correlation functions. Although the charge excitation gap remains finite in the thermodynamic limit ($\Delta_{\rm c}\left(L=\infty\right)/t=0.95 \pm 0.05$), the spin excitation gap remains negligible within the statistical error bar of our method ($\Delta_{\rm s}\left(L=\infty\right)/t \approx 0.03 \pm 0.03$ (see Fig.~\ref{fig:gaps}). 

\section{VI. Summary}
Employing the $Z_2$ slave-spin, VMC, and QMC approaches we studied the phase diagram of the Kitaev-Hubbard model 
as a prototype model for some of the layered materials such as iridates and $\alpha$-RuCl$_3$, where the magnetic ions with strong correlations are sitting on the vertices of a honeycomb lattice. In particular, we provided evidences for a spin liquid phase for $1 \lesssim t^{\prime}/t \lesssim 1.11$ in the phase space for $U/t = 5$. The observation of a quantum spin liquid state in this model may contribute to our understanding of materials like $\alpha$-RuCl$_3$, shown to be proximate to the Kitaev spin liquid.

\section{acknowledgments}
\begin{acknowledgments}
The authors gratefully acknowledge the Iran Science Elites Federation (Saramadan) for funding this project.
\end{acknowledgments}

\section{APPENDIX A: Details of the slave-spin formulation of the Kitaev-Hubbard model}

We employ the $\mathbb{Z}_2$ slave-spin approach to examine the phase diagram of the Kitaev-Hubbard model :

\begin{equation}\label{H_KH1}
H_{\mathrm{KHu}} =  - \sum_{\langle ij \rangle_\gamma} \left \lbrace c_i^{\dagger}\frac{(t+t^{\prime}\sigma_{\gamma})}{2}c_j + \mathrm{h.c.} \right \rbrace + U \sum_j \left(n_{j,\uparrow}-\frac{1}{2}\right)\left(n_{j,\downarrow}-\frac{1}{2}\right).
\end{equation}

{Notably, in contrast to other slave-spin methods~\cite{Kotliar:PRL1986, Florens:PRB2002, Florens:PRB2004, deMedici:PRB2005,mardani2011slave,HassanPRB2010}, this approach involves a discrete $\mathbb{Z}_2$ gauge fluctutions and extend the local Hilbert space with dim=4 into a space with dim=8.} 
Within this method, each fermion is represented using an auxiliary spin as follows:
\begin{equation}\label{slavespin1}
c_{i,s}^{(\dagger)}=2I^x_i f_{i,s}^{(\dagger)},
\end{equation} 
where, $I_i^x$ is a pseudo-spin carrying the electron charge property, and $f_{i,s}$ is the annihilation operator for a pseudo-fermion. The above representation expands the local physical $4$-dimensional Hilbert space into an $8$-dimensional one. However, the physical subspace is identified by imposing these constraints: (i) on the spinon sector, $f_{i,s}^\dag f_{i,s} = c_{i,s}^\dag c_{i,s}$, and (ii) on the charge sector, $I_z=+1/2$ for the doubly occupied as well as the empty electron states and $I_z=-1/2$ for the singly occupied electron states. Hence, $I_x$ switches the local electron number parity, namely $(-1)^{n_i}$, where $n_i = n_{i,\uparrow}+n_{i,\downarrow}$. 

In this representation the Kitaev-Hubbard model \eqref{H_KH1} is given by the following Hamiltonian

\begin{equation}\label{KHM_slave_spin1}
H = 4 \sum_{\langle ij \rangle_\gamma} I_i^x I_j^x f_{i,\alpha}^{\dagger}t_{\alpha,\beta}^{\gamma}f_{j,\beta} + \frac{U}{2} \sum_j I_j^z.
\end{equation}

In this representation, the Hubbard interaction term has been transformed into a simple form involving the field $I_z$. However, the hopping term becomes quartic, which can be further simplified using the mean-field approximation. This facilitates an exact examination of the Hubbard term and an approximate analysis of the hopping term.  

\section{APPENDIX B: Phase diagram of the model}\label{phase_diagram}
In this section, we apply the mean-field approximation to Eq. \eqref{KHM_slave_spin1} to obtain phase diagram. At the mean-field level, this Hamiltonian takes the form
\begin{equation}\label{MFH}
H_{\rm MF} = \sum_{\langle ij \rangle_\gamma} g_{ij} (f_{i,s}^{\dagger}t_{s,s'}^{\gamma}f_{j,s'} + \text{h.c.}) -4 \sum_{\langle ij \rangle_\gamma} J_{i,j} I_i^x I_j^x + \frac{U}{2} \sum_j I_j^z.
\end{equation}
where the coefficients are defined as
\begin{equation}\label{mean_fields1}
\begin{split}
&g_{i,j}=4\langle I_i^x I_j^x\rangle, \quad J_{i,j}=-\langle f_{i,s}^{\dagger} t_{s,s'}^{\gamma} f_{j,s'}^{\dagger} \rangle + \text{c.c.}
\end{split}
\end{equation}
and $t^\gamma_{s,s'} = \frac{1}{2}\left(t\delta_{s,s'} + t'\sigma_{s,s'}^\gamma\right)$. In this scenario, the total Hamiltonian decouples into the sum of a hopping Hamiltonian of fermions and a transverse-field Ising model. Accordingly, the eigenstates of the Hamiltonian can be expressed as the product of the eigenstates of these two sectors: $|\psi \rangle = |\psi_f\rangle | \psi_I \rangle$ which breaks the local U(1) symmetry of the system down to its $\mathbb{Z}_2$ subgroup \cite{Ruegg:PRB2010}.

Since $J_{i,j}$ varies along the three different neighboring directions and is spin-dependent, we introduce a translational invariant mean-field ansatz with  parameters $J_{\gamma}$ and $g_{\gamma}$, where $\gamma = x,y,z$, each corresponds to a bond direction.  
For the Ising model, we employ a 6-site cluster mean-field approximation to calculate $g_{\gamma}$ parameters.
Minimizing the energy for various ansatz states, we obtain the phase diagram shown in Fig.\ref{phase_diagram}(a). 

\begin{figure}[t]
\centering
\includegraphics[width=0.7\linewidth]{phase_diagram.pdf}
\caption{
{(a) $\mathbb{Z}_2$ slave-spin phase diagram of the Kitaev-Hubbard model. At strong $U/t$ the Mott phases are VBS (AFI), striped VBS (i-Neel) and quantum spin liquid (QSL) phases. (b) An illustration of Mott phases: thick green bonds correspond to finite $g_{\gamma}$, while $g_{\gamma}=0$ on the thin bonds. (c) Magnetic phase diagram of the Kitaev-Hubbard model. (d) Transverse static susceptibility around the $\Gamma$ point, considering different $t'$ values, corresponding to circles panel (c), with $U=3.5$. The inset graph displays the value of $\chi_{+-}^0(\mathbf{q})$ for $t' = 0.9$ around the $\Gamma$ point. The right-hand inset also illustrates the first Brillouin zone.}}\label{phase_diagram1}
\end{figure}

The transverse Ising field Hamiltonian illustrates two distinct phases. By increasing the transverse field strength ($U$) beyond a critical value $U_c\left(t'/t\right)$, the system undergoes a quantum phase transition from a ferromagnetic to a paramagnetic state. The paramagnetic phase is characterized by $\langle I_i^x\rangle=0$. Since $I_x$ encapsulates the charge degrees of freedom by construction, in this phase the charge fluctuations are suppressed. Thus, the Mott phase is characterized by vanishing the quasi-particle weight
$Z = 4 \langle I_i^x \rangle^2$ \cite{Florens2004, HassanPRB2010}. As shown Fig.~\ref{phase_diagram1}(a), the model is in the semimetal (SM) phase in the limit of weak coupling. The Mott phase arises for large values of $U/t$. It is seen that the system becomes less correlated as $t'/t$ increases, i.e., at the larger values of spin-orbit coupling, a stronger $U$ is required for the Mott transition to occur. 

The various Mott phases are distinguished from each other by noting that the values of $g_{ij}$ preserve or break the lattice symmetries. Since $g_{ij}= m/m^*$ is inversely proportional to effective mass of quasi-particles, ansatz with $g_{ij}=0$ on some bonds breakes the lattice symmetries. Note that not all couplings could be zero; otherwise the spinon Hamiltonian vanishes, which is not physical. 

We found that by increasing $t'/t$, three distinct Mott phases arise in the Kitaev-Hubbard model: valence bond solid (VBS), striped VBS (S - VBS), and the quantum spin liquid (QSL) phases as shown Fig.\ref{phase_diagram1}(a). A schematic representation of the coupling $g_{ij}$ in different Mott phases is demonstrated in Fig.\ref{phase_diagram1}(b). In VBS phase the coupling $g_{\gamma}$ becomes non-zero only for one of the bonds, for instance $g_{x}\neq0$ and $g_{y}=g_{z}=0$. 
{The corresponding ansatz breaks the rotational symmetry, and the spinons form resonant states on the bond. Despite the fact that the time-reversal symmetry is broken in the original Hamiltonian, our mean-field construction does not give magnetically ordered states. However, given that for large values of $U$ and small $t'$, the system is effectively described by an antiferromagnetic Heisenberg model (see Equation (2) of the main text), the resonant VBS phase corresponds to an antiferromagnetically ordered state. As the parameter $t'$ increases, the mean-field solution yields an ansatz with $g_{x}=g_{y}\neq0$ and $g_{z}=0$. Hence, the spinons can only disperse along the stripe paths (S - VBS) where spinons disperse unidirectionally as shown in Fig.\ref{phase_diagram1}(b). The phase should also be understood as a state with spin density wave ground state. These two magnetic phases will be further explored in the subsection \ref{magnetic_instability}.}

A lattice-symmetry preserving phase with $g_{x}=g_{y}=g_{z}\neq0$ arises in the limit of strong $U/t$ and $t'/t$. According to Equation (2) of the main text, the magnetic phase with $t'=t$ corresponds to the KSL phase. Our results however show that the latter phase survives and remains robust for a wide range of $U$ and $t'$, see the light blue region in Fig.\ref{phase_diagram1}(a). In this phase, the system is described by the fermionic sector of the mean-field Hamiltonian \eqref{MFH} yielding four energy bands with Dirac points. Hence, this phase represents a gapless spin liquid. 

\subsection{Magnetically ordered phases and magnetic fluctuations}\label{magnetic_instability}
To investigate the possible magnetically ordered phases in the KHu model, we express the Hubbard interaction as $H_U = \sum_j (-\frac{U}{6}\sigma_j^2+ \frac{U}{4})$ where $\sigma_j=c_{j,\alpha}^{\dagger}\sigma_{\alpha,\beta}c_{j,\beta}$. The interaction is decoupled using the Hubbard-Stratonovich transformation by taking the antiferromagnetic order parameter as $M = \frac{U}{6}\langle (\sigma_{j,A}^z- \sigma_{j,B}^z)\rangle$, where $A$ and $B$ are sublattices. As shown in Fig.\ref{phase_diagram1}(c). The system lacks any magnetic order for small values of $t'$ and $U$. By increasing $U$, the system undergoes a transition into an antiferromagnetic (AF) phase.

We further explore the AF phase by examining the spin fluctuations. The magnetic fluctuations within the random phase approximation (RPA) are described as $\langle \sigma_{\bold{q}}^{\mu} \sigma_{\bold{-q}}^{\nu} \rangle = \frac{1}{I^2} \lbrace [I_m^{-1} - \chi_0(\bold{q})]^{-1} - I_m \rbrace_{\mu,\nu}$, where, $\chi_0(\bold{q})$ denotes the bare susceptibility, $I=U/6$, $I_m=\mathrm{diag}(I, 2I, 2I)$, $\sigma_{\bold{q}}^{\mu} = \sigma_{\bold{q},A}^{\mu} -\sigma_{\bold{q},B}^{\mu}$. 

In Fig.\ref{phase_diagram1} (d) we plot the transverse static magnetic susceptibility for different values of $t'$ (marked by circles in panel (a)) at fixed $U=3.5$. For $t'=0$ the sharp peak at the $\Gamma$ point is associated to AF magnetic ordering. By increasing $t'$ the peak intensity is decreased and eventually disappears at $t'=1$ indicating the complete melting of the long-range order. Note that at relatively large values of $t'$ the transverse susceptibility developes two broad peaks around the $\Gamma$ point indicating that at intermediate values of $t'$ an incommensurate antiferromagnetic (i-Neel) phase \cite{Liang:PRB2014} sets in. Though the $\mathbb{Z}_2$ slave-spin mean-field phases have lower energy, the obtained lattice-symmetry broken phases using this approach could correspond to AF and  i-Neel phases, and the supression of ordered phases at large $t'$ is an indication of a spin liquid phase preserving the symmetries.

\section{APPENDIX C: $\mathbb{Z}_2$ gauge theory of the Kitaev-Hubbard model}
In this section, we investigate the fluctuations around the mean-field approximation of the Kitaev-Hubbard model, represented by Eq. \eqref{H_KH1}, using the $\mathbb{Z}_2$ slave-spin framework. Since the mean-field Hamiltonian, Eq. \eqref{MFH}, explicitly breaks the $\mathbb{Z}_2$ symmetry, we introduce fluctuating fields $\sigma^z_{ij}$ on the spatial links to restore the original $\mathbb{Z}_2$ symmetry. Consequently, the mean-field Hamiltonian can be expressed in the following reformulated form 

\begin{equation}\label{MFH_2}
H_{\rm MF} = \sum_{\langle ij \rangle_\gamma} g_{ij}\sigma^z_{ij} (f_{i,s}^{\dagger}t_{s,s'}^{\gamma}f_{j,s'} + \text{h.c.}) -4 \sum_{\langle ij \rangle_\gamma} J_{ij} \sigma^z_{ij} I_i^x I_j^x + \frac{U}{2} \sum_j I_j^z.
\end{equation}
The Hamiltonian remains invariant under the following $\mathbb{Z}_2$ gauge transformation \cite{Fiete:PRL2012}
\begin{equation}
u_i = (-1)^{\left(f_{i,\alpha}^{\dagger} f_{i,\alpha} + I_i^z - \frac{1}{2}\right)} \prod_{j(i)} \sigma_{ij}^x.
\end{equation}
Here, $j(i)$ denotes the nearest neighbors of site $i$. The projection operator onto the physical subspace is given by $P = \prod_i P_i$, where $P_i = \frac{1}{2} \left(1 + u_i \right)$. We can express the projection operator for the new physical subspace using a time-dependent fluctuating field as follows \cite{Senthil:PRB2000}
\begin{equation}
P_i = \prod_i \frac{1}{2} \sum_{\sigma_{i \tau}^z=\pm1} e^{i\pi/2(1-\sigma_{i,\tau}^z)\left(f_{i,\alpha}^{\dagger}f_{i,\alpha}+I_i^z-\frac{1}{2}\prod_{j(i)}\sigma_{ij}^x
\right)}.
\end{equation}
In the subsequent analysis, we apply the path integral formalism to derive the $\mathbb{Z}_2$ gauge theory for this model. The partition function within the physical subspace is expressed as
\begin{equation}
Z = \mathrm{Tr}(e^{-\beta H}P).
\end{equation}

\begin{widetext}
In terms of fermionic coherent states, the formulation takes the following form
\begin{equation}
Z = \int \prod_i d\bar{f}_{i,\alpha}df_{i,\alpha} \sum_{\lbrace I^x_i \rbrace} \sum_{\lbrace \sigma^z_{ij} \rbrace} e^{-\bar{f}_{i,\alpha}f_{i,\alpha}} \langle -\bar{f}_{i,\alpha};I^x_i;\sigma^z_{ij}|(e^{-\epsilon H}P)^M| f_{i,\alpha};I^x_i;\sigma^z_{ij} \rangle
\end{equation}
Here, $\epsilon = \beta/M$, and $f_{i,\alpha}$ and $\bar{f}_{i,\alpha}$ are Grassmann variables defined as
\begin{equation}
\begin{split}
& |f_{i,\alpha}\rangle = e^{-f_{i,\alpha}\hat{f}_{i,\alpha}^{\dagger}}|0\rangle,\\
& \langle \bar{f}_{i,\alpha}|=\langle 0| e^{\bar{f}_{i,\alpha}\hat{f}_{i,\alpha}}.
\end{split}
\end{equation}
Furthermore, we have integrated over all configurations of the pseudospin variables and the fluctuating fields $\sigma_{ij}^z$, where
\begin{equation}
\begin{split}
&I^x_i |I^x_i\rangle = I^x_i |I^x_i\rangle \\
&\sigma^z_{ij} |\sigma^z_{ij}\rangle = \sigma^z_{ij} |\sigma^z_{ij}\rangle.
\end{split}
\end{equation}
In the following, we introduce identity operators for the Grassmann variables, pseudospins, and $\mathbb{Z}_2$ fields, $\sigma_{ij}^z$, in different time bases
\begin{equation}
\begin{split}
&\int d\bar{f}_{i,\alpha,\tau} d\bar{f}_{i,\alpha,\tau} e^{-\bar{f}_{i,\alpha,\tau}f_{i,\alpha,\tau}}|f_{i,\alpha,\tau}\rangle \langle \bar{f}_{i,\alpha,\tau}|=1,\\
&\sum_{\lbrace I_{i,\tau}^x \rbrace} |I_{i,\tau}^x \rangle \langle I_{i,\tau}^x|=1,\\
&\sum_{\lbrace \sigma_{ij,\tau}^z \rbrace} |\sigma_{ij,\tau}^z \rangle \langle \sigma_{ij,\tau}^z|=1.
\end{split}
\end{equation}
This allows us to express the partition function as
\begin{equation}
Z = \int \prod^M_{\tau=1} \prod_{i,\alpha} d\bar{f}_{i,\alpha,\tau} df_{i,\alpha,\tau} \sum_{\lbrace I^x_{i,\tau} \rbrace} e^{-\sum^M_{\tau=1}\sum_{i,\alpha}\bar{f}_{i,\alpha,\tau} f_{i,\alpha,\tau}} \langle \bar{f}_{\tau}, I^x_{\tau}|e^{-\epsilon H} P | f_{\tau-1},I^x_{\tau-1}\rangle.
\end{equation}
Utilizing the relationships
\begin{equation}
\begin{split}
&\langle \bar{f}_{\alpha,\tau} | e^{i\pi/2(1-\sigma^z_{i \tau})\hat{f}_{i,\alpha,\tau}^{\dagger}\hat{f}_{i,\alpha,\tau-1}}|f_{i,\alpha,\tau-1}\rangle = e^{\sigma^z_{i\tau}\bar{f}_{i,\alpha,\tau}f_{\alpha,\tau-1}},\\
& \langle I^x_{\tau}|e^{i\pi/2(1-\sigma^z_{i\tau})(I_{i\tau}^z-1/2)} \left(\sum_{\lbrace I^z_{i\tau} \rbrace} |I^z_{\tau}\rangle \langle I^z_{\tau}| \right) | I^x_{\tau-1} \rangle=e^{i\pi/2 \left(I_{i\tau}^z-1/2\right)(2(I^x_{i\tau-1} - I^x_{i\tau}) + (1-\sigma^z_{i\tau}))}\\
& \langle \sigma_{ij}^z(\tau) | e^{i\pi/4(1-\sigma_{i\tau}^z)(1-\prod_{j(i)}\sigma_{ij}^x)}| \sigma_{ij}^z(\tau) \rangle = \sum_{\lbrace \sigma_{ij}^x(\tau)\rbrace} e^{i\pi/4(1-\sigma_{i,\tau}^z)(1-\prod_{j(i)}\sigma_{ij}^x)} e^{i\pi/4(1-\sigma_{ij}^x)(\sigma_{ij}^z(\tau-1)-\sigma_{ij}^z(\tau))},
\end{split}
\end{equation}
and introducing the variable transformation
\begin{equation}
\sigma_{i\tau}^z f_{i,\tau-1,\alpha} \rightarrow f_{i,\tau,\alpha},
\end{equation}
the partition function takes the form
\begin{equation}
\begin{split}
Z = &\prod_{\tau=1}^M \prod_{i,\alpha} d\bar{f}_{i,\alpha,\tau} df_{i,\alpha,\tau} \sum_{\lbrace I^x_{i,\tau} \rbrace} \sum_{\lbrace I^z_{i,\tau} \rbrace} \sum_{\lbrace \sigma^z_{i,\tau} \rbrace} \sum_{\lbrace \sigma^x_{i,\tau} \rbrace} e^{-S},
\end{split}
\end{equation}
where the total action is expressed as
\begin{equation}
S = S^f + S^{\mathrm{Ising}} + S^{\sigma} + \epsilon \sum_{\tau=1}^M H(I^x_{\tau}, I^z_{\tau}, \bar{f}_{\tau}, f_{\tau-1}),
\end{equation}
and the individual actions are given by
\begin{equation}
\begin{split}
&S^f = \sum_{\tau=1}^M \sum_{i,\alpha} \bar{f}_{i,\alpha,\tau}(\sigma_{i,\tau}^z f_{i,\alpha,\tau-1}- f_{i,\alpha,\tau})\\
&S^{\mathrm{Ising}} = -\frac{i\pi}{2}\sum_{\tau=1}^M \sum_{i} \left(I_{i,\tau}^z-1/2\right)(2(I^x_{i,\tau-1} - I^x_{i,\tau}) + (1-\sigma^z_{i,\tau}))\\
&S^{\sigma} =  -\frac{i\pi}{4}\sum_{\tau=1}^M \sum_{i}\left[(1-\sigma_{i,\tau}^z)(1-\prod_{j(i)}\sigma_{ij}^x(\tau)) + (1-\sigma_{ij}^x(\tau))(\sigma_{ij}^z(\tau-1)-\sigma_{ij}^z(\tau))\right].
\end{split}
\end{equation}
The complete mean-field Hamiltonian is expressed as
\begin{equation}
H_{\rm MF} = \sum_{\langle ij \rangle_\gamma} g_{ij} (f_{i,s}^{\dagger}t_{s,s'}^{\gamma}f_{j,s'} + \text{h.c.}) -4 \sum_{\langle ij \rangle_\gamma} J_{i,j} I_i^x I_j^x + \frac{U}{2} \sum_j I_j^z.
\end{equation}
By considering fluctuations around the mean-field and neglecting domain fluctuations, the Hamiltonian assumes the form given in equation \eqref{MFH_2}. This expression can be decomposed into the sum of two actions, $S_{\mathrm{MF}}$ and $S_U$, as follows
\begin{equation}
\begin{split}
&S_{\mathrm{MF}} = -\epsilon \left(\sum_{\langle ij \rangle_\gamma} g_{ij} \sigma_{ij}^z (f_{i,s}^{\dagger}t_{s,s'}^{\gamma}f_{j,s'} + \text{h.c.}) -4 \sum_{\langle ij \rangle_\gamma} J_{i,j} \sigma_{ij}^z I_i^x I_j^x \right),  \\
&S_U = -\epsilon \frac{U}{2} \sum_j I_j^z.
\end{split}
\end{equation}
To express our final theory exclusively in terms of $f$, $\bar{f}$, $I^x$, and the gauge fields, we perform the sum over all configurations of $I_i^z$
\begin{equation}
\sum_{\lbrace I_i^z \rbrace} e^{-\left(S_U+S^{\mathrm{Ising}}\right)} = e^{2\ln \coth(\epsilon U)} \sum_i I^x_i \sigma_{i,i-\tau}^z I^x_{i-\tau} e^{-S_{B}}.
\end{equation}
Here, $S_{B}$ denotes the Berry phase term, defined as
\begin{equation}
e^{-S_B}=\prod_{i} \sigma_{i,i-\tau}^z.
\end{equation}
Considering the translational invariance, we express $J_{ij}$ in terms of $J_{\gamma}$ and $g_{ij}$ in terms of $g_{\gamma}$, where 
$\gamma = x,y,z$. Therefore, the pseudospin part of the action becomes
Considering translational invariance, we express $J_{ij}$ in terms of $J_{\gamma}$ and $g_{ij}$ in terms of $g_{\gamma}$, where $\gamma = x, y, z$. Consequently, the pseudospin component of the action takes the form
\begin{equation}
S_{I_x} = 4 \epsilon \sum_{\langle ij \rangle_\gamma} J_{\gamma} I_i^x I_j^x - 2\ln \coth(\epsilon U) \sum_i I^x_i \sigma_{i,i-\tau}^z I^x_{i-\tau}.
\end{equation}
Given that $\epsilon$ is a small parameter, we determine its value by equating the expressions $\kappa_{\gamma} \equiv 4\epsilon J_{\gamma}$ and $2 \ln \coth(\epsilon U)$. This allows us to rewrite the total action in the spacetime lattice basis (i.e., in terms of $i$ and $j$). Consequently, the total partition function becomes
\begin{equation}\label{partition_function}
\begin{split}
&Z = \int D \bar{f}_{i,\alpha} D f_{i,\alpha} \sum_{\lbrace I^x_{i} \rbrace} \sum_{\lbrace \sigma^z_{ij} \rbrace} \sum_{\lbrace \sigma^x_{ij} \rbrace} e^{-S_{Z_2}\lbrace \bar{f}, f, I_x, \sigma_z, \sigma_x\rbrace}\\
&S_{Z_2}= S_{I_x} + S_f + S_{\sigma} + S_B\\
&S_{Ix} = -\sum_{\langle ij \rangle_{\gamma}} \kappa_{\gamma} I_i^x \sigma_{ij}^z I_j^x\\
&S_f = -\sum_{ij} A_{\alpha,\beta}^{\gamma} \bar{f}_{i,\alpha} \sigma_{ij}^z f_{j,\beta},
\end{split}
\end{equation}
where $A_{\alpha,\beta}^{\gamma} = \epsilon B_{\alpha,\beta}^{\gamma}$. Here, $B_{\alpha,\beta}^{\gamma}$ takes the values $g_{\gamma} t_{\alpha,\beta}^{\gamma}$ on spatial links and $-\delta_{i,j} \delta_{\alpha,\beta}/\epsilon$ on temporal links. By summing over all configurations of $\sigma^z_{ij}$ in the action $S_{\sigma}$, the resulting expression assumes the following form
\begin{equation}
\sum_{\lbrace \sigma^z_{ij} \rbrace} e^{-S_{\sigma}} = \sum_{\lbrace \sigma^z_{ij} \rbrace} e^{\frac{i\pi}{4}\sum_{\tau=1}^M \sum_{i}\left[(\sigma_{i,\tau}^z-1)(\prod_{j(i)}\sigma_{ij}^x(\tau)-1) + (\sigma_{ij}^x(\tau)-1)(\sigma_{ij}^z(\tau)-\sigma_{ij}^z(\tau-1))\right]} \propto e^{ G \sum_i \prod_{j(i)}\sigma_{ij}^x + I \sum_{\langle ij \rangle} \sigma_{ij}^x}.
\end{equation}
Given that spinons within the Mott phase manifest gapless behavior, in contrast to the gapped characteristics of pseudospin excitations, a low-energy theory describing spinons and gauge fields emerges from integrating out the degrees of freedom associated with pseudospin excitations. This process is mathematically expressed as
\begin{equation}
\sum_{\lbrace I^x_{i} \rbrace} e^{\sum_{\langle ij \rangle_{\gamma}} \kappa_{\gamma} I_i^x \sigma_{ij}^z I_j^x} \propto e^{K \sum_{\hexagon}\prod_{\langle ij\rangle \in \hexagon}\sigma_{ij}^z}  
\end{equation}
Here, the parameters $K$, $G$, and $I$ are strictly non-negative. Consequently, the effective Hamiltonian governing the gauge fields takes the following form \cite{Fiete:PRL2012}:
\begin{equation}
H_{\sigma} = -K  \sum_{\hexagon}\prod_{\langle ij\rangle \in \hexagon}\sigma_{ij}^z - G \sum_i \prod_{j(i)}\sigma_{ij}^x - I \sum_{\langle ij \rangle} \sigma_{ij}^x.
\end{equation}
\end{widetext}

\section{APPENDIX D: Effective Hamiltonian of the System in the Quantum Spin Liquid Phase}
As previously mentioned, the mean-field Hamiltonian for spinons remains gapless in the KSL phase, while the effective Hamiltonian for charges acquires a gap. Consequently, the low-energy Hamiltonian of the system can be expressed in terms of spinons as
\begin{equation}
H_f = \sum_{\langle ij \rangle_\gamma} g_{\gamma}(f_{i,s}^{\dagger}t_{s,s'}^{\gamma}f_{j,s'} + \text{h.c.}) = \sum_{\langle ij \rangle_\gamma} U_{ij}^{\gamma}.
\end{equation}
where $g_{\gamma}$ is non-zero in all directions and each site is characterized by $n_i = 1$ within this Mott phase. Therefore, the pseudofermions appear as charge-neutral spinons, which can be expressed in terms of spin operators as $S_i^{\gamma} = f_{i,\alpha}^{\dagger} \sigma_{\alpha,\beta}^{\gamma} f_{i,\beta}$. Any operator acting within this subspace must preserve the condition $n_i = 1$. The minimal operator commutes with $n_i = 1$ is $(U_{ij}^{\gamma})^2$, which operates along the $\gamma$-direction and can be expressed as $(U_{ij}^{\gamma})^2 = g_{\gamma}^2 S_i^{\gamma} S_j^{\gamma}$ in term of spin operators. Consequently, the Hamiltonian governing the low-energy excitations in the KSL phase is given by
\begin{equation}
H_{\rm KSL} = - K \sum_{\langle ij \rangle_\gamma}S_{i}^{\gamma}S_{j}^{\gamma},
\end{equation}
Here, the parameter $K$ represents a positive constant. This Hamiltonian precisely mirrors the Kitaev model, confirming the existence of a gapless spin liquid state in this phase.

\section{APPENDIX E: Mean-Field Theory of Magnetism in the Kitaev-Hubbard Model}
\begin{figure}[h!]
\centering
\includegraphics[width=0.7\linewidth]{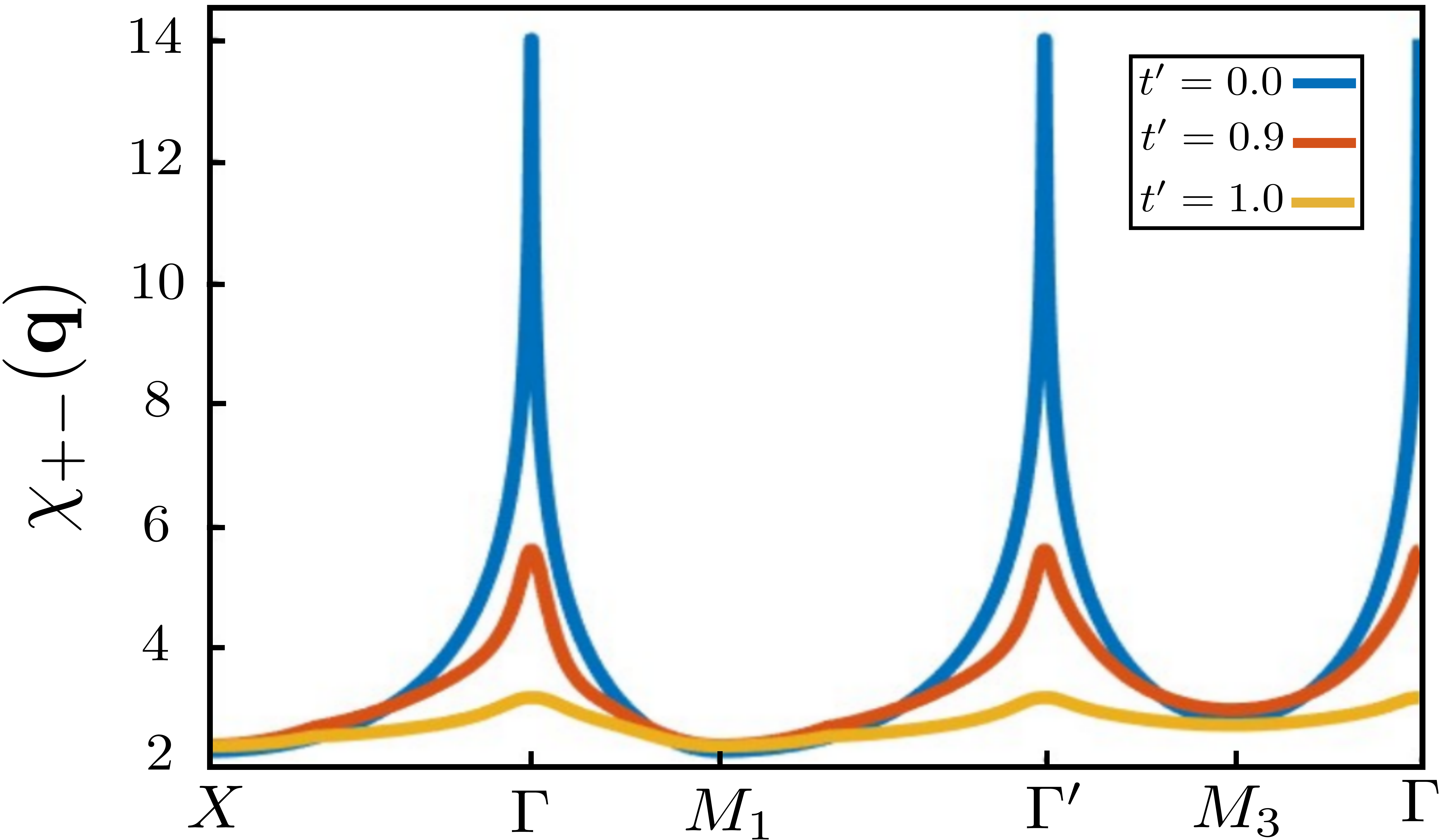}
\caption{Transverse static susceptibility for various values of $t'$ with $U = 3.5$}\label{chi_pm}
\end{figure}

To explore the magnetic phases within the Kitaev-Hubbard model, we employ the path integral method. The non-interacting component of the system's Hamiltonian in momentum space is expressed as
\begin{equation}
H_0 = c_{\bold{k}}^{\dagger}H_{\bold{k}}  c_{\bold{k}} = \sum_{\bold{k}} 
\left( \begin{matrix} c_{\bold{k},A}^{\dagger} & c_{\bold{k},B}^{\dagger}\end{matrix}\right)
\left( \begin{matrix}
0 & T_{\bold{k}} \\
T_{\bold{k}}^{\dagger} & 0
\end{matrix}\right)
\left( \begin{matrix} c_{\bold{k},A} \\ c_{\bold{k},B}\end{matrix}\right).
\end{equation}
Here, $T_{\bold{k}}$ is defined as $T_{\bold{k}} = e^{-i\bold{k} \cdot a_1} t_x + e^{-i\bold{k} \cdot a_2} t_y + t_z$, where $\bold{a}_1 = a \left( \frac{3}{2}, \frac{\sqrt{3}}{2} \right)$ and $\bold{a}_2 = a \left( \frac{3}{2}, -\frac{\sqrt{3}}{2} \right)$. Moreover, the interaction term can be expressed in terms of spin operators as
\begin{equation}
H_U = \sum_j \left( -\frac{U}{2}(\sigma_j^z)^2+ \frac{U}{4}\right),
\end{equation}
where $\sigma_j^z = (n_{j, \uparrow} - n_{j, \downarrow})$ denotes the magnetic fluctuations around the $z$-axis. However, this approach might overestimate $\sigma_j^z$, as magnetic fluctuations can occur along all three coordinate axes. To properly account for fluctuations in all directions, $H_U$ can be written as
\begin{equation}
H_U = \sum_j \left(-\frac{U}{6}\sigma_j^2+ \frac{U}{4}\right),
\end{equation}
where $\boldsymbol{\sigma}_j = c_{j, \alpha}^{\dagger} \boldsymbol{\sigma}_{\alpha, \beta} c_{j, \beta}$. We then use the Hubbard-Stratonovich transformation to reframe the interaction term in terms of a fluctuating Weiss field. The Weiss field $\boldsymbol{\phi}_j$ is expressed as
\begin{equation}
\bold{m}_j = \boldsymbol{\phi}_j - \frac{U}{3} c_{j,\alpha}^{\dagger} \boldsymbol{\sigma}_{\alpha,\beta}c_{j,\beta}.
\end{equation}
where $\bold{m}_j$ denotes a white noise. In this context, the total action of the system is given by
\begin{equation}
S[\bar{c}, c, \phi]=\int_0^\beta d\tau \Big [ \sum_{\bold{k}} \bold{c}_\bold{k}^{\dagger}(\partial_\tau + H_\bold{k})\bold{c}_{\bold{k}} - \sum_j \boldsymbol{\phi}_j \cdot (\boldsymbol{\sigma}_{j,A} + \boldsymbol{\sigma}_{j,B}) + \frac{3}{2U}\sum_j {\boldsymbol{\phi}}^2_j  \Big].
\end{equation}
Next, we decompose the fields $\boldsymbol{\phi}_j$ into two new fields that represent the ferromagnetic and antiferromagnetic orders within each unit cell
\begin{equation}
\begin{split}
&\boldsymbol{\phi}_{\rm AF, j} = \frac{1}{2}\left( \boldsymbol{\phi}_{j,A} - \boldsymbol{\phi}_{j,B}\right),\\
& \boldsymbol{\phi}_{\rm F,j} = \frac{1}{2}\left(\boldsymbol{\phi}_{j,A} + \boldsymbol{\phi}_{j,B} \right),
\end{split}
\end{equation}
where $A$ and $B$ denote the two sublattices of the honeycomb lattice. The action in terms of these new fields is given by
\begin{widetext}
\begin{equation}
\begin{split}
S[\bar{c}, c, \boldsymbol{\phi}_{\rm AF}, \boldsymbol{\phi}_{\rm F}] = \int_0^\beta d\tau \Big [& \sum_{\bold{k}} \bold{c}_{\bold{k}}^{\dagger}(\partial_\tau + H_{\bold{k}})\bold{c}_{\bold{k}} - \sum_j \boldsymbol{\phi}_{\rm F,j} \cdot (\boldsymbol{\sigma}_{j,A} + \boldsymbol{\sigma}_{j,B}) + \frac{3}{U}\sum_j {\boldsymbol{\phi}}^2_{\rm F,j} \\ & - \sum_j \boldsymbol{\phi}_{\rm AF,j} \cdot (\boldsymbol{\sigma}_{j,A} - \boldsymbol{\sigma}_{j,B}) + \frac{3}{U}\sum_j \boldsymbol{\phi}_{\rm AF,j}^2 \Big].
\end{split}
\end{equation}
\end{widetext}
For large values of $U$ and $t' = 0$, the effective spin model of the system corresponds to a Heisenberg model with a ground state that exhibits antiferromagnetic order. Therefore, the saddle point approximation using an antiferromagnetic ansatz appears to provide an appropriate effective field for small values of $t'$. By minimizing the action with respect to the uniform mean fields $\boldsymbol{\phi}_{\mathrm{F},j} = \boldsymbol{\phi}_{\mathrm{F}} = 0$ and $\boldsymbol{\phi}_{\mathrm{AF},j} = \boldsymbol{\phi}_{\mathrm{AF}} = M\hat{z}$, the antiferromagnetic order parameter is determined to be
\begin{equation}
M = \frac{U}{6}\langle \sigma_{j,A}^z- \sigma_{j,B}^z\rangle.
\end{equation}
This antiferromagnetic order can, in principle, be nonzero along all three coordinate axes, although for simplicity, we consider it along the $z$-axis. The RPA magnetic fluctuations around the mean-field are described by
\begin{equation}
\langle \sigma_{\bold{q}}^{\mu} \sigma_{\bold{-q}}^{\nu} \rangle = \frac{1}{I^2}\Bigg \lbrace \Big[I_m^{-1} - \chi_0(\bold{q})\Big]^{-1} - I_m \Bigg \rbrace_{\mu,\nu},
\end{equation}
where $\chi_0(\bold{q})$ denotes the bare susceptibility, $I = U/6$, $I_m = \mathrm{diag}(I, 2I, 2I)$, and $\sigma_{\bold{q}}^{\mu} = \sigma_{\bold{q},A}^{\mu} - \sigma_{\bold{q},B}^{\mu}$. The static transverse response is shown in Fig.~\ref{chi_pm}.


\bibliographystyle{apsrev4-2}
\bibliography{Refs}

\end{document}